# Open data hackathon as a tool for increased engagement of Generation Z: to hack or not to hack?

Anastasija Nikiforova [0000-0002-0532-3488]

University of Tartu, Institute of Computer Science, Narva mnt 18, 51009 Tartu, Estonia

nikiforova.anastasija@gmail.com

**Abstract.** A hackathon is known as a form of civic innovation in which participants representing citizens can point out existing problems or social needs and propose a solution. Given the high social, technical, and economic potential of open government data (OGD), the concept of open data hackathons is becoming popular around the world. This concept has become popular in Latvia with the annual hackathons organised for a specific cluster of citizens – Generation Z. Contrary to the general opinion, the organizer suggests that the main goal of open data hackathons to raise an awareness of OGD has been achieved, and there has been a debate about the need to continue them. This study presents the latest findings on the role of open data hackathons and the benefits that they can bring to both the society, participants, and government. First, a systematic literature review is carried out to establish a knowledge base. Then, empirical research of 4 case studies of open data hackathons for Generation Z participants held between 2018 and 2021 in Latvia is conducted to understand which ideas dominated and what were the main results of these events for the OGD initiative. It demonstrates that, despite the widespread belief that young people are indifferent to current societal and natural problems, the ideas developed correspond to current situation and are aimed at solving them, revealing aspects for improvement in both the provision of data, infrastructure, culture, and government- related areas that will allow moving towards innovative knowledge-based society and a smart city.

**Keywords:** Civic Innovation, Hackathon, Generation Z, Government, Knowledge-based society, Open Data, OGD, Open Government Data, Social Innovation, Society 5.0, Sustainability.

# Introduction

Open data hackathons are considered a creative and "out-of-the-box" approach to civic innovation, also known as "social innovation", described by an unconventional way of thinking and an openness to create solutions in a new and creative way [1-2], which are becoming increasingly popular all over the world. The specificity and perhaps the main value of these government-induced open data engagement initiatives [2] is that they bring together people with different backgrounds, experience, knowledge, skills, and expertise [1,3] in one place for short periods of time, thus supporting intense bursts of creativity [4] to create services or solutions that will benefit the public by creating a *win-win* scenario for all participants [5]. The diverse skills and knowledge of participants facilitate and promote innovation and learning as participants generate and evaluate ideas from different perspectives [3]. The hackathon provides a solution-oriented environment for the co-creation (collaborative creation) of social services, free from hierarchical constraints. Another important, but not the only, goal of such events is to raise an awareness of the open government data (OGD) among citizens, making them the main input that should be used to create solutions or services within the hackathon, thus indicating their value and unlocking their potential. The definition of open data hackathons before pandemic also referred to the mode in which they were held - an offline, face-to-face competition [2], which changed due to the COVID-19 pandemic, when hackathons turned into a virtual space, and remain to be online events. Citizen engagement in these hackathons can greatly contribute to the uptake and adoption of OGD as a whole [6]. In addition, as suggested in [5,7], organizing hackathons or digital innovation competitions is an opportunity for aspiring entrepreneurs to collaborate with external partners, promote new ideas and launch a new start-up. Therefore, in the hope of promoting digital services, which can be a fundamental element in the development of new business or start-ups, or other type of benefit, government agencies are coordinating such events [5].

Another popular trend is the organization of hackathons for participants representing Generation Z (Gen Z), which is and should be an interesting target audience, considered to have the best digital capabilities [8] being so-called "*digital natives*", because they have never experienced life before the Internet [9], which, along with social media, became part of their daily lives and socialization [10]. This makes Gen Z to be unique, since no other generation has lived in the era when technology is so readily available at such a young age [9]. In the context of Gen Z, a recent study [11] inspected OGD as a governance strategy as a potential factor in promoting millennial and Generation Z trust in government institutions. They found that OGD has a positive impact on millennials and Gen Z satisfaction with public outcomes and trust in institutions. In this way, the authors encourage civil servants to implement open data strategies to improve youth attraction and involvement in democratic institutions. Open data hackathons can be an excellent opportunity to be used. This is even more so considering motivation factors for participating in hackathons defined in [6], where 4 of 5 factors - intrinsic motivation, extrinsic motivation, effort expectancy, social influence, and data quality- are considered valid by default for Gen Z, where only the "data quality" is not met by default but can be seen as a test item that Gen Z is likely to be able to evaluate. Therefore, in line with the argument proposed in [6] - citizen engagement is one step further than the use of OGD, this study suggests that Gen Z engagement may be the next towards revealing and solving societal issues and pointing out challenges related to OGD.

The OGD hackathon for Gen Z participants has been held annually in Latvia since 2018, but the organizers believe that the goal of hackathons to raise an awareness of OGD has been achieved, and the debate about the need for their further organization has been ongoing on for the past two years. This study aims to highlight the need for Gen Z hackathons as a source of feedback, identifying opportunities for improvements, and seeking ideas for the development and maintenance of a sustainable and citizen-oriented smart city and *knowledge-based society 5.0*. The latter - *Society 5.0* or super smart society also referred to as society of imagination - is seen as the next from of the society defined as "*imagination society, where digital transformation combines with the creativity of diverse people to bring about "problem solving" and "value creation" that lead us to sustainable development*" as defined in [12] and is characterized by 5 key areas: (1) *problem solving and value creation*, (2) *diversity*, (3) *decentralization*, (4) *resilience*, (5) *sustainability and environmental harmony*. It is also expected to contribute to the achievement of the Sustainable Development Goals (SDG) adopted by the United Nations, both sharing the same objectives.

To meet the objective of the study, this paper will first establish the role of open data hackathons by examining current research on the topic. It aims to demonstrate various perspectives, including the benefits that open data hackathons can bring. This is done by carrying out systematic literature review (SLR). Then, empirical research of 4 case studies of open data hackathons for Generation Z participants held between 2018 and 2021 in Latvia is conducted to understand which ideas dominated and what were the main results of these events for the OGD initiative, and whether the COVID-19 led to a decrease in the diversity of solutions developed expected by the organizers. It demonstrates that, despite the widespread belief that young people (in Latvia) are indifferent to current societal and natural problems, the ideas developed correspond to current problems and are aimed at solving them, revealing aspects for improvement in both the provision of data, infrastructure, culture, and government related areas that will allow moving towards a sustainable, resilient, and innovative knowledge-based society or *Society 5.0* and a smart city. The analysis is conducted by the hackathon mentor, which should affect positively the interpretability of the results. The results clearly indicate the need for further organization as a source of feedback from the most modern cluster of society, identifying opportunities for improvements and seeking ideas for a more coherent and robust understanding of how open data crisis-management can be established and maintained. This idea was originally presented in [13].

The paper is organized as follows: Section 2 provides an overview of the existing body of knowledge with reference to a SLR and its results, Section 3 provides an overview of Gen Z open data hackathons in Latvia, analyses ideas developed in 5 editions of this hackathon and observations made by participants on the OGD initiative, as well as the observations made by mentor, while Section 4 concludes the paper.

## Related Research

To identify relevant literature for study, forming the knowledge base, a systematic literature review (SLR) was conducted. This was done by searching digital libraries covered by Scopus and Web of Science (WoS) that index well-known publishers of peer-reviewed literature such as ScienceDirect (Elsevier), Springer, Emerald, ACM, IEEE, Taylor & Francis, Sage, Wiley-Blackwell, Oxford University Press, Cambridge University Press etc. These databases were queried for keywords (1) ""*hackathon" AND ("Open Government Data" OR "open data" OR "OGD") AND ("Generation Z" OR "Gen Z)"*, (2) ""*hackathon" AND ("Open Government Data" OR "open data" OR "OGD")"*, (3) *"("Open Government Data" OR "open data" OR "OGD") AND ("Generation Z" OR "Gen Z)"*. These search terms were applied to the article title, keywords, and abstract to limit the number of papers to those, where these objects were primary research object rather than mentioned in the body, e.g., as a future work. Only articles in English were considered, while in terms of scope, both journal articles, conference papers, and chapters were studied. The first search returned no results, i.e., there are no studies on the open data hackathons discussed in the context of Gen Z. The $2^{nd}$ query resulted in 38 articles in Scopus and 5 in WoS. After comparing the resulting sets and eliminating duplicates, 39 studies remained for their further examination covering either open data hackathons or open data in the context of the Gen Z. The third query resulted in one article - [11] referred to in the Introduction. In terms of the research area, these studies tend to be different, i.e. in Scopus, most studies belong to Computer Science (24 of 38), Social Sciences (14), Business, Management and Accounting (6), Engineering (5), Mathematics (5), and Decision Sciences (4), while in WoS - Computer Science (3 of 5), Public Administration (2), Geography (1), Government Law (1), Information Science Library Science (1). These studies are diverse in nature - from testing an idea during a hackathon as an experiential setting representing different domains in which to gather feedback, exploring trends in a specific area such as developing software or a strategy to set up a start-up, to conceptualizing the motivation for participating in hackathon etc. They mostly indicate the high and varied value of hackathons both for the organizers, i.e., for the government enabling *government as a model of an open data activist*" [6], participants, and for society. Let discuss them in more detail.

In [5] the authors examine goals and design strategies that contribute to the successful execution of open data hackathons to understand the co-ordination between the multiple stakeholders and improve the execution of open data hackathons and innovation competitions. The outcomes indicate that the most critical design strategy was the involvement of mentors in the event and the level of support provided to nascent entrepreneurs to accelerate their creativity, develop applications, and launch their prototypes on the market. The authors argue that prior studies have not thoroughly concentrated on the planning and assessment processes of these competitions, stressing that scholars have not compared the execution of different strategies in many hackathons or digital innovation competitions rather focusing on the actions that the organizers of a specific hackathon have implemented or the challenges they have faced.

Yuan & Gasco-Hernandez [14] address a gap in research that refers to open innovation outcomes and contribution to public value creation by referring to open data hackathons in US. It defines the concept of open innovation as the leverage of external resources and knowledge provided by citizens or other stakeholders for innovation to help solve public problems, which contributes to the creation of public value. They found that open data hackathons promote *procedural rationality*, also called *procedural legitimacy*, *democratic accountability*, and *substantive outcomes*. They also emphasize that *citizen engagement* and *participation*, as well as *network building* are more important than physical products or solutions as the outcomes of open innovation initiatives.

Jaskiewicz et al. [15] investigated the opportunities of leveraging a hackathon format to empower citizens by increasing abilities to use open data to improve their neighbourhoods and communities. The discussion presented is based on five case studies of civic hackathons organized in five European cities. The research revealed specialized learning and collaborative alignment as two mutually complementary aspects of the learning processes involved, which were achieved with the help of high-fidelity and low-fidelity prototypes, respectively. The study identified three key factors needed to sustain social learning ecosystems beyond the hackathon events, and with the purpose of democratizing smart city services. These factors include (a) *supporting individuals in obtaining specific expert knowledge and skills*, (b) *building/nurturing data-literate activist communities of practice made up of citizens with complementary set of expert skills*, and (c) *enabling members of these communities to prototype open-data services*. The twofold purpose originally formulated for civic hackathons considered (O4C) was to support citizens in discovering new opportunities for meaningful open data applications, and *bottom-up*, *community-driven learning* and *sharing of data literacy skills*. These goals are closely interlinked. This allowed them to develop a general framework that promises to serve as a guideline for defining relevant key performance (KPI) and behavioural indicators to assess learning in future civic hackathons. The framework considers (a) *individual learning*, (b) *community capacity*, and (c) *learning through prototyping* as three mutually enforcing learning activities within civic hackathons. However, study by Molinari [16] argues that initiatives such as the O4C better articulate some of the psychosocial dimensions that explain the resistance to data disclosure, especially on the part of public administration, the impact of political culture on the management of public goods, the aims and motivations of the stakeholders involved, guided by different interests and competencies, and the role of outside advocacy groups.

Carroll & Beck [17] demonstrated that hackathons provide insights into practices and goals, where water quality was the subject of their interest, i.e., diverse local initiatives for water quality testing and threat mitigation were identified during the hackathons, allowing problems to be framed and brainstorm session to be organized to collectively discuss possible solutions. This allowed them to introduce, or rather co-design, the concept of *platform collectivism* as an alternative to *platform capitalism*, concluding that the gathering and sharing community watershed data is an inherently collective endeavour that co-produces community engagement and water security

Purwanto et al. [2] explored the motivations of citizens to be engaged in hackathons by conducting a case-study of open data hackathons that represent to one particular domain – agriculture or farming. They found out that a list of factors motivate citizens to participate in OGD hackathons constituting a framework of citizens' motivation to engage in open data hackathons: (1) *intrinsic motivation*, represented by constructs such as fun and enjoyment and intellectual challenge, (2) *extrinsic motivation*, constituted by performance expectancy or relative advantage and learning and developing skills, (3) *effort expectancy* related to the ease of use, (4) *social influence* related to influence from a social relationship and contributing to societal benefits, and (5) *data quality* with the reference to accuracy. With the reference to this study, it should be noted that 4 of 5 factors describing motivation can be considered valid by default for Gen Z, where only the "data quality" defined as "*data that are fit for use by data consumers*" is not met by default but can be seen as a test item that Gen Z is likely to be able to evaluate.

Gama [18] argues that hackathons, while being a very promising idea, tend to suffer from several problems, such as (1) *functional requirements defined for applications in civic hackathons identified based on needs, interests or experience of developers instead of societal needs*, (2) *concerns about the quality of the software*, i.e. the quality of the developed service or solution, (3) *poor organization and task management*, (4) *issues related to release and maintenance of the developed solution*. The investigation was conducted by surveying participants in three civic hackathons. The 3rd assumption was not proven with a partly rejection of the fourth assumption, if the maintenance of the developed item was the case for the vast majority of participants and at least version control tools were employed, however, as regards the release, it really turned out that most of the developed prototypes were not finished and, in many cases, abandoned, especially those that are not recognized as winners. Bad or low quality of the developed software was also rejected, as the authors saw traces of design and testing practices. However, there was evidence to support claims that civic apps are based on the experience of developers rather than the actual needs of citizen, where roughly half of the respondents indicated that the requirements for their apps were based only on their experience. This result, however, cannot be generalized, as 40% of the participants in two of three hackathons studied did additional (online) research.

Mainka et al. [19] discussed the role of hackathons in the development of mobile applications (*m-apps*) using OGD collectively, intended to solve urban problems, as it is not known whether the output of hackathons leads to value-added city services, as well as challenges that governments face in making open data available. The latter, i.e., the challenges of making OGD available can be seen as both (1) *a political challenge* where politicians are afraid of losing their monopoly in public affairs, (2) *a legal challenge* associated with *security*, *privacy* and *copyright* as key arguments of "protecting" data from the public, (3) *governance challenge* – motivation to open data and the need to collaborate with business and citizens, (4) *human resource challenge* related to the variety of skills required to prepare data suitable for reuse, which is seen as a complex task, (5) *IT infrastructure challenge*, making online services available to the public, and (6) *IT budget challenge*, with the reference to cases where the government implements charging models for their data. Their investigation showed that most hackathons gather people, who are programmers or designers, which is not in line with the cities / urban open data vision, i.e., most citizens do not feel affiliated to join hackathons. It is still unclear whether hackathons are really capable of changing something and involving citizens in governmental processes, although people come together as a community and develop public services, which can help make the whole city and its residents smarter.

There are also studies confirming the power and usefulness of intensive team-oriented hackathons for the development of previously unseen services or solutions, such as Data & Analytics Framework (DAF) [20], Biomedical Data Translator program ("Translator") [21], a planning tool for farmers that compiles crop calendars, agroclimatic data, and historical records of production enabling geospatial navigation of agricultural activities and crowdsourced updates of calendars ("CropPlanning") [22]. They also make it possible to define requirements for further developments, as was the case for a system called "Theophrastus" [23] designed to assist biologists in their research on species and biodiversity, which supports automatic annotation of documents through entity mining, and provides services using Linked Open Data (LOD) in real-time.

To sum up, while the topic of hackathon is relatively popular in the literature with some studies devoted to open data hackathons and datathons, conducted SLR indicates a scientific gap in the knowledge on the Gen Z role in them making this study unique.

## Gen Z open data hackathon

This section is devoted to Gen Z open data hackathons in Latvia. First, it provides a brief overview of the general structure of hackathons. It then provides an overview of the ideas developed between 2018 and 2021 , using a qualitative approach to analyse the collected data. It then discusses an overview of issues identified by participants related to open data, infrastructure etc., data on which were collected through interviews with participants during the hackathon during the post-pitch Q&A session.

Open data hackathons for representatives of Gen Z are held annually in Latvia, where the first edition took place in 2018. They are organized by the Latvian Association of Open Technologies in cooperation with governmental / public agencies, including the entity responsible for the OGD in Latvia, namely the Latvian Ministry of Environmental Protection and Regional Development (VARAM) as part of the jury or organizer, depending on the edition. Their main idea, as defined by organizers but also found in the literature [24], is to raise awareness of OGD, thereby facilitating an increase in their use by this cluster of citizens representing the future and development of the country.

Compared to the traditional hackathon model, including the model referred to in the previous section, these hackathons are longer, i.e., the participants are given several weeks to make their solutions and services as polished as possible. This is also the case because the hackathon takes place in the autumn during the fall semester, which tends to negatively impact participation rates if the traditional hackathons model is chosen. This form of hackathons, being closer to virtual hackathons is generally found to be a good way of promoting inclusiveness and digital participative collaboration [25]. This also allowed the organizers to switch to the fully virtual / online mode in 2020 (due to pandemic) relatively easy. In short, participants apply for a hackathon, then they get together to develop their idea, gather teammates (if not done before), brainstorm, get some knowledge from the hand-outs and workshops delivered by the organizers, mentors, partner etc., and then present their ideas at a very high-level during the first pitch session. This pitch session allows participant to receive initial feedback from mentors, organizers, and other participants, as well as to ask their questions on the sketched idea. This also allows the organizers to recommend or assign mentors to teams depending on the idea and the nature of the questions (theoretical or practical, technical etc.). Then, teams work independently on their own outside of the event, organizing their work independently and communicating with the mentors, who guide them. Then two or three weeks later (depending on the edition and requests for the extension), the second pitch session is organized, when progress, i.e., the mock-up or prototype is demonstrated and elaborated on by participants, allowing mentors to see "from idea to solution" progress and sustainability of the service, including its significance and value for the economy, society etc. After the pitch session, a team of mentors selects the most promising solutions and services to be presented at the hackathon final – usually 10 teams, which bring together a wider audience, including, but not limited to, industry representatives. They are sometimes specially invited by the organizers and mentors, depending on the nature of the developed services to get the attention of relevant stakeholders, thereby trying to prevent the idea from being abandoned or discarded, which is a common problem of hackathons [5,26-27], and increasing chances to get the support – technological, financial etc. Participants are provided with another portion of feedback on their solutions and / or services, where mentors share their experience and suggestions, and also allow them to choose the right direction for their ideas and suggestions for monetizing them. And then, one week later, the final pitch session takes place, where selected services and solutions are presented, considering previously received feedback, where the final decision is made on the winners both at the level of hackathon and supporters, where the latter can be public agencies or business representatives. All in all, these hackathons last approximately 1 month.

**Ideas developed over 5 editions**

This section discusses ideas developed in Gen Z open data hackathons between 2018 and 2021 in Latvia. This is done using a qualitative approach analysing the data collected by the organizers of the hackathon, including the author, who is the mentor of this hackathon. Thus, this study provides an analysis made from the mentor perspective with its direct involvement / participation, which allows a more accurate interpretation of the data collected and the observations made.

*Open data hackathon of 2018*

The year 2018 was characterized by the predominance of solutions related to mobility and transport aimed at simplifying route planning, bicycle parking - finding a spot or requesting / voting for organizing a new one. However, planning of leisure time for both tourists (by means of crowdsourcing) and local citizens, and recommenders for choosing of place of living depending on the location and level of crime, and the determination of a kindergarten with the shortest queues for a child, were also popular. While most of them are something fairly typical for open data-based solutions regardless of the country, although contributing to its sustainable development or improvement of daily routines (being also in line with most SDG), the latter – the determination of a kindergarten with the least queues was of very important for young parents in that year, where there was an expressed issue with finding kindergarten and very limited information about the process making it very non-transparent.

*Open data hackathon of 2019*

In 2019, the question of how to choose the best place to live based on the nearest infrastructure and prices was popular, while solutions different from 2018 were an open data-based tool for a unified assessment of Latvian schools and a forecast of their future development being especially relevant for Latvia that year, given poor results shown in examinations and the upcoming education reform. Last but not least, there was a solution to simplify the search for a medical institution and corresponding doctors for patients with a focus on rare diseases, which was a very timely solutions given the digitization of medical services and the development of national e-health system, which at first raised a lot of concerns among citizens.

*Open data hackathon of 2020*

The 2020 hackathon was the first edition that took place in an online mode six months after the start of the pandemic. Although it was assumed that all solutions will be related to the fight against the pandemic and crisis management, this was not the only case considered by the participants. Moreover, COVID-19 was more seen as factor to be considered in solutions to allow leisure planning mostly for citizens, but for some teams, for tourists, with the reference on how to recover the tourism during and after pandemic (to identify the least crowded areas, places, or routes), since the tourism is of high economic value for Latvia. There was an expressed popularity of solutions that allow young people to choose a university or profession based on current statistics on their salaries, employment etc. Some ideas were influenced by the call of the organizers to test the data prepared for their further publication on the national OGD portal (after the feedback collected about them) – data of the Register of Enterprises of the Republic of Latvia. Some ideas were oriented on preventing climate change, with two more solutions dedicated to agriculture. All in all, only two solutions focused on the COVID-19 only - interactive maps based on the current rate of a spread of the disease and the consequences by regions.

*Open data hackathon of 2021*

The 2021 hackathon was defined by the organizers as an open geospatial data hackathon, encouraging participants to use the geodata available. This edition dominated by solutions focused on regional development and urban planning, with a focus on identifying problem areas to be addressed in an appropriate regional development plan. Some of the proposed solutions refers to the gamification design, or at least the use of game elements, which is also in line with current trends set by both theoreticians and practitioners [28]. One solution was dedicated to waste management, with 4 more floods and fire forecasts and their risk assessment, air quality and pollution monitoring, with another for planning bicycle routes considering the landscape and infrastructure appropriateness. Likewise in 2020, only 1 solution was purely COVID-19 related, suggesting the creation of a single source for crawling, and presenting all information related to COVID-19, with one more considering its consequences - internet coverage outside the capital of Latvia for those who work from other regions with much more limited access to the internet (if any). The solution mentioned above – a single source of information related to the pandemic, would gather all informative materials about the current situation that would be provided to users in a visual manner, supplementing these data and relevant visualizations with forecasts / predictions of risk to be affected by COVID-19 in various regions, which should be achieved by means of using neural networks. In addition, the team planned a built-in chatbot and the opportunity to get in touch with users of the service, sending out the most critical notifications about changes in the regions of interest. Although the idea of this solution overlaps with the solutions currently available, it was presented before the actual creation of such a website in Latvia, thereby indicating the interest of citizens (even so young) in solutions of this nature, especially given that the idea presented is more advanced compared to the introduced. The latter was also recognized by the developers of the above solution. This demonstrates the potential of open (government) data and is in line with the potential of the OGD to transform the society into a Society 5.0 - society of imagination.

**Key observations reported by participants**

The above discussed structure of hackathons allowed to gather a feedback or observations made by participants while working with the open data and the portal on which they are accessible. These data were collected through interviews with participants during the hackathon during a post-pitch Q&A session. Unfortunately, this feedback was collected as a result of unstructured interviews during short Q&A sessions without the possibility of further communication and collecting more detailed feedback, which should definitely be the case of hackathons as this is a great opportunity to get an opinion of someone who actually used the data from their locating, downloading, discovering, distilling, scrutinizing, refining, with their further transformation into a prototype, including the design and development of such, and has more likely encountered challenges that are valid for open data and the open data portal (also in line with [6]).

Nevertheless, the feedback collected is in line with expert assessment of the Latvian OGD and portal [29], being also in line with evidence found in the literature [5]. More precisely, the main critical feedback or the issues identified by the participants are:

(1) *poor data quality*, where both the quality of the metadata and the data, i.e., the content of the data (completeness, accuracy, credibility), were referred to;

(2) *outdated data* with the reference to the issues related to the *timeliness* of data and *data recency*, which tend to be violated;

(3) data in some datasets tend to be *poorly structured*;

(4) the *lack of "valuable data"* found to be the most frequently reported issue;

(5) the *inability to use data through API*.

Data quality and data credibility / reliability that was a topic of interest for decades, has become more topical in the context of OGD, which tend to suffer from low quality. At the same time, there is limited research on open data quality with reference to the content of datasets with a higher interest in metadata quality or defining open data quality as the compliance with open data principles, lacking a data management and data quality management perspective. The rise in the popularity of this topic was observed during pandemics, when data on COVID-19 became a very hot topic for both researchers, practitioners, including small and medium-sized business (SME) and citizens. This proves again that although the availability of data is a prerequisite, a shift "*from quantity to quality*" should be made. This is especially important given that it is not always clear who is responsible for this – the owners of the data portal, government, the publishers of the data, or data user, who should analyze data quality analysis prior their use.

For the lack of "valuable data", commonly referred to as "high value data", the 2021 hackathon provided evidence that not only the imagination of citizens or users and the desire to create value form the OGD for society matters, but also the diversity of data. However, not only the content of the data matters here, but also the readiness of the country under consideration for digitization at all levels. It has been shown that in the case of Latvia, especially the latter – the level of digitization – is in some cases insufficient for the development of some crisis management services. More precisely, one of the ideas was to develop an application that would allow citizens to plan the use of public transport, tracking not only its timetable/ schedule and current location (whether it arrives on time or is delayed), but also the number of people in it to avoid overcrowding. This would allow an assessment of the need to travel at a specific time and vehicle when transport is overcrowded and the ability to choose another vehicle or time to reach their destination in a safer way, thus allowing both individuals and the general public to be protected from maintenance in an overcrowded vehicle. On the one hand, there are open data available on the OGD portal, where registrations of an e-ticket are fixed, which is partially in line with the objective, but they are not provided in real-time, which significantly reduces their value and contradicts with the modern trends regarding OGD, because only approximate data can be obtained, by analyzing historical data, instead of real-time data. In addition, although they would allow to fix the number of passengers who have just entered the transport in question, these data still do not provide actual data on the number of passengers, since passengers who have exited the transport are not recorded. However, this could be addressed by either equipping transport with sensors to monitor the number of passengers entering and leaving, thereby allowing for real-time and accurate data that would create value, or by other advanced solutions such as computer vision, as suggested participants of the hackathon. It should be mentioned that the question of the need to integrate sensors into transport was discussed in Latvia before the pandemic, but it was decided not to implement them. The crisis has highlighted the emerging need of smart solutions, including urban data, and the need to follow current trends. It also points on the complexity of the OGD ecosystem, which cannot be separated from other areas of national development.

The lack of API pointed to in earlier editions of the hackathons (2018-2019) seems to be at least partly resolved, which proves the current state of the OGD. More precisely, the owners of the OGD portal have ensured that all datasets provided in *.csv* format are automatically supplemented with the ability to retrieve them through the API.

All in all, hackathons indicate or assign some prerequisites to data, according to which the data available must comply with principles such as machine-readability, timeliness, and data currency, requiring regular updates of the data, high data quality and data relevance being a valuable asset for business and society, as well as these data should be easy to retrieve, i.e., API support becomes almost a prerequisite.

Although the feedback was collected through unstructured interviews with no opportunity to collect more detailed feedback, the observations made by the participants are consistent with expert assessment, suggesting that the hackathon can be a valuable asset for the government and public agencies to get rich and relevant feedback on the current state and identify opportunities for improvement, which can sometimes be accompanied by recommendations. The value of these also lies in the fact that this information comes from real data users making it possible to look at it from citizen's lens.

**Key implications from hackathons: a mentor's perspective**

Although Generation Z is associated with the term "*digital natives*", the point to be considered, however, is the age of the current representatives of Gen Z, which is relatively low and, as a result, some limitations arise from it. In other words, some of them do not hold a secondary school diploma, which automatically makes them less prepared for the classical hackathon's paradigm, where participants come together for hours or days to develop solutions in many cases having relevant skills and experience – being in line with findings that in most cases participants of hackathons tend to be developers or software engineers. This requires some efforts on the part of mentors and organizers, as well as the organization of practical workshops, which, however, although provides support and some basic knowledge, cannot substitute a real experience.

This leads to the fact that some ideas are not implemented or remain at the level of a sketch or mock-up, while there are many – those presented in the final, which are already working prototypes. It should be emphasized that the exclusion of the developed ideas depending on the level of their implementation should not be decisive, where the maturity of the general idea is of higher importance, which is especially true considering the age of the participants. In other words, according to the existing body of knowledge, if an idea is mature enough and proves its societal or business value, it deserves consideration even in final, when it can be presented for a wider audience and help participants find supporters – technological, business etc., who will help to implement it and monetize, thereby revealing the full potential of the idea and hackathons in general. Otherwise, the risk of abandoning the idea, and project discontinued is high as proves both the experience and literature [18]. This is also in line with [24], according to which the key take-away of hackathons should not only awareness raising or physical products as outcomes [14], where *citizen engagement* and *networking* are most relevant. Latvian hackathons have also proven to be events that allow businesses and government *to see gaps in the current state-of-the-art of both social, technical, and sociotechnical nature*, and get ideas on how they can be addressed presented by participants representing the most modern part of society, with relevant ideas that are one step ahead of the ideas proposed within entities – businesses or government. The question is *whether the representatives of these entities are open to the proposed solutions* or rather resistant to these changes and to the realization of the reality regarding the problems identified.

Regarding the prototypes developed during the hackathons, although it can be speculated that they are be developed not only by the participants, but also by other people not participating in the hackathon for the sake of the needed competence, i.e., parents, students etc., most participants demonstrate their expertise during pitch sessions and progress discussions with mentors, providing some evidence that the Gen Z are digital natives. These participants are characterized not only by the high level of knowledge and skills showed in the development of their prototypes, but also by the acquisition of new knowledge and skills in a very short terms, e.g., when they are suggested to turn to another technology. Nonetheless, there are some gaps in their knowledge, such as version control of their solutions or (an accurate) licencing of published source code, that older participants more likely have – although even this is argument is sometimes posed under question [18]. This together with the current body of knowledge about Gen Z and specificities of learning methods they require, which should be significantly different from those used for previous generations born in less technologically developed world, being integrated with various games, mobile apps [10], allows to define the proposition that *a hackathon can also serve as a part of the educational process* [13].

## Conclusions

In promoting the use of open data, governments sought to be seen not (only) as "*government as a platform*", but also as "*a government as an open data activist model*", in which the government not only provides an open data infrastructure but also promotes its use by citizens, the private sector, or the government itself [6], often organize activities in the form of a hackathon contests, where citizens (and businesses) compete to pitch an idea or the design for a service [30]. However, the role of these hackathons is not always obvious. Many see them only as a form of raising awareness of OGD and the corresponding portal, and once this goal seems to be achieved, there tend to be a debate about the need to continue their organization. This is also the case for Latvia and open data hackathons for Generation Z seen as the future of our society.

The aim of this paper was to highlight the need for open data hackathons and hackathons for Gen Z as the most modern part of the society, including their role as a source of feedback, identifying opportunities for improvement, and seeking ideas for sustainable and citizen-centric development and maintenance of the smart city and knowledge-based society. To achieve this goal, this study first conducts a SLR, thereby building a knowledge base of what is currently known about open data hackathons, in particular, in the context of Gen Z, known as *digital natives*. It was revealed that although the topic of hackathon is relatively popular in the literature, and young people make up the majority of the participants, with some studies devoted to open data hackathons, there is a scientific gap in knowledge, including the role of Gen Z in (open) data hackathons.

Then, an empirical study of 4 open data hackathons for Generation Z organized in Latvia was carried out to understand which ideas dominated and what were the main results of these events for the OGD initiative. An analysis of the developed ideas, solutions, and services, i.e., regardless of the stage of output maturity, whether it is a mock-up or a working prototype, indicates that, despite the widespread opinion about the indifference of young people to current societal and natural problems, developed ideas correspond to the current situation and relevant problems for the country aimed at solving them. This study also confirms the results presented in [19], i.e., while it remains unclear whether hackathons are truly capable of making a difference and engaging citizens in government processes, it is clear is that people come together as a community and develop or co-create governmental services, which also contributes to the digital and open data literacy of participants. This can help make the whole city and its residents smarter, as well as identify and suggest improvement to public services even if there is little open data, although they should be sufficiently structured, of high quality [31] and represent high-value data, also potentially capable to contribute to the development of a value-creating, sustainable open data ecosystem [32]. It also allows to suggest that this form of citizen engagement facilitates the shift or transition from the *Information Society*, known as *Society 4.0*, to *Society 5.0*, or what Verhulst et al. [33] call "*Collective Intelligence*", also known as "*wisdom-of-crowd*" [34].

This study rejects the organizer's assumption that the COVID-19 pandemic will lead to a decrease in the diversity of solutions, where the pandemic was seen as a factor to be considered when presenting different solutions and services, including crisis management services. The results clearly point to the need for further organization as a source of feedback from the most modern cluster of society. In other words, they should not be organized to raise awareness of open data alone, where not only citizen engagement and network building or physical products or solutions as outcomes of open innovation initiatives [14,24] are important, but also the ability to identify opportunities for improvements in the provision of open data, infrastructure, culture, and government areas that will allow and facilitate movement towards a sustainable, resilient, innovative knowledge-based society and a smart city, seeking for a more robust understanding of how open data crisis management can be established and maintained.

# References

Toros, K., Kangro, K., Lepik, K. L., Bugarszki, Z., Sindi, I., Saia, K., & Medar, M. *Co-creation of social services on the example of social hackathon: The case of Estonia*. International Social Work, 0020872820904130 (2020)

Purwanto, A., Zuiderwijk, A., & Janssen, M. Citizens' motivations for engaging in open data hackathons. In *International Conference on Electronic Participation* (pp. 130-141). Springer, Cham (2019)

Pe-Than, E. P. P., & Herbsleb, J. D. Understanding hackathons for science: Collaboration, affordances, and outcomes. In *International Conference on Information* (pp. 27-37). Springer, Cham. (2019)

Taylor, N., & Clarke, L. *Everybody's hacking: Participation and the mainstreaming of hackathons*. In CHI 2018 (pp. 1-2). Association for Computing Machinery (2018)

Kitsios, F., & Kamariotou, M. Digital innovation and entrepreneurship transformation through open data hackathons: Design strategies for successful start-up settings. *International Journal of Information Management*, 102472 (2022)

Purwanto, A., Zuiderwijk, A., & Janssen, M. *Group development stages in open government data engagement initiatives: a comparative case studies analysis*. In International Conference on Electronic Government (pp. 48-59). Springer, Cham (2018)

Nolte, A. Touched by the Hackathon: a study on the connection between Hackathon participants and start-up founders. In *2nd ACM SIGSOFT International Workshop on Software-Intensive Business: Start-ups, Platforms, and Ecosystems* (pp. 31-36) (2019)

Basantes-Andrade, A., Cabezas-González, M., & Casillas-Martín, S. *Digital competences relationship between gender and generation of university professors*. International Journal on Advanced Science, Engineering and Information Technology, 10(1), 205-211 (2020)

Prensky, M. Digital natives, digital immigrants part 2: Do they really think differently? *On the horizon* (2001)

Szymkowiak, A., Melović, B., Dabić, M., Jeganathan, K., & Kundi, G. S. *Information technology and Gen Z: The role of teachers, the internet, and technology in the education of young people*. Technology in Society, 65 (2021)

Gonzálvez-Gallego, N., & Nieto-Torrejón, L. Can open data increase younger generations' trust in democratic institutions? A study in the European Union. *Plos one*, *16*(1), (2021)

World Economic Forum. Modern society has reached its limits. Society 5.0 will liberate us (2019), online, available at: https://www.weforum.org/agenda/2019/01/modern-society-has-reached-its-limits-society-5-0-will-liberate-us/, last accessed 10.06.2022

Nikiforova, A. Gen Z open data hackathon - civic innovation with digital natives: to hack or not to hack? In *2022 Ongoing Research, Practitioners, Posters, Workshops, and Projects of the International Conference, EGOV-CeDEM-ePart 2022*. CEUR-WS (2022) (in print)

Yuan, Q., & Gasco-Hernandez, M. Open innovation in the public sector: creating public value through civic hackathons. *Public Management Review*, *23*(4), 523-544 (2021)

Jaskiewicz, T., Mulder, I., Morelli, N., & Pedersen, J. S. Hacking the hackathon format to empower citizens in outsmarting" smart" cities. *IxD&A*, *43*, 8-29 (2019)

Molinari, F., Concilio G. Culture, Motivation and Advocacy: Relevance of Psycho Social Aspects in Public Data Disclosure. In *Proceedings of the 17th European Conference on Digital Government* (pp. 86-95). Academic Conferences and Publishing International (2017)

Carroll, J., & Beck, J. Co-designing platform collectivism. *CoDesign*, *15*(3), 272-287 (2019)

Gama, K. Preliminary findings on software engineering practices in civic hackathons. In *2017 IEEE/ACM 4th International Workshop on CrowdSourcing in Software Engineering* (pp. 14-20). IEEE (2017)

Mainka, A., Hartmann, S., Meschede, C., & Stock, W. G. Open government: Transforming data into value-added city services. In *Citizen's Right to the Digital City* (pp. 199-214). Springer, Singapore (2015)

Fallucchi, F., Petito, M., & Luca, E. W. D. Analysing and visualising open data within the data and analytics framework. In *Research Conference on Metadata and Semantics Research* (pp. 135-146). Springer, Cham (2018)

Fecho, K., Ahalt, S. C., Arunachalam, S., Champion, J., ... & Biomedical Data Translator Consortium. Sex, obesity, diabetes, and exposure to particulate matter among patients with severe asthma: Scientific insights from a comparative analysis of open clinical data sources during a five-day hackathon. *Journal of biomedical informatics*, *100*, 103325 (2019)

Grajales, D. F. P., Mejia, F., Mosquera, G. J. A., Piedrahita, L. C., & Basurto, C. Crop-planning, making smarter agriculture with climate data. In *2015 Fourth International Conference on Agro-Geoinformatics (Agro-geoinformatics)* (pp. 240-244). IEEE (2015)

Fafalios, P., & Papadakos, P. Theophrastus: On demand and real-time automatic annotation and exploration of (web) documents using open linked data. *Journal of web semantics*, *29*, 31-38 (2014)

Macedo Guimarães, L., Bitencourt, R. S., Chrusciak, C. B., Derenevich, M. G., Poncini, C. R., Okumura, M, & Junior, O. C. Sustainability Hackathon: Integrating Academia and Companies for Finding Solutions for Socio-environmental Problems. In *Integrating Social Responsibility and Sustainable Development* (pp. 591-607) Springer, Cham. (2021)

Charvat, K., Bye, B. L., Kubickova, H., Zampati, F., Löytty, T., Odhiambo, K., ... & Kamau, W. (Capacity Development and Collaboration for Sustainable African Agriculture: Amplification of Impact Through Hackathons. *Data Science Journal*, *20*(1) (2021)

Ayele, W. Y., Juell-Skielse, G., Hjalmarsson, A., Johannesson, P., & Rudmark, D. Evaluating Open Data Innovation: A Measurement Model for Digital Innovation Contests. In *PACIS* (p. 204) (2015)

Frey, F. J., & Luks, M. The innovation-driven hackathon: one means for accelerating innovation. In *21st European Conference on Pattern Languages of Programs* (pp. 1-11) (2016)

Simonofski, A., Zuiderwijk, A., Clarinval, A., & Hammedi, W. Tailoring open government data portals for lay citizens: A gamification theory approach. *International Journal of Information Management*, *65*, 102511 (2022)

Nikiforova, A., & Lnenicka, M. *A multi-perspective knowledge-driven approach for analysis of the demand side of the Open Government Data portal*. Government Information Quarterly, 38(4), 101622 (2021)

Sieber, R. E., & Johnson, P. A. *Civic open data at a crossroads: Dominant models and current challenges*. Government information quarterly, 32(3), 308-315 (2015)

Matheus, R., Ribeiro, M. M., & Vaz, J. C. Brazil towards government 2.0: Strategies for adopting open government data in national and subnational governments. In *Case Studies in e-Government 2.0* (pp. 121-138). Springer, Cham (2015)

Van Loenen, B., Zuiderwijk, A., Vancauwenberghe, G.,... & Flores, C. C. (2021). Towards value-creating and sustainable open data ecosystems: A comparative case study and a research agenda. *Je-DEM-eJournal of eDemocracy and Open Government*, *13*(2), 1-27.

Verhulst, S., Addo, P. M., Young, A., Zahuranec, A. J., Baumann, D., & McMurren, J. Emerging Uses of Technology for Development: A New Intelligence Paradigm. Available at SSRN 3937649 (2021)

Suran, S., Pattanaik, V., Kurvers, R., Antonia Hallin, C., De Liddo, A., Krimmer, R., & Draheim, D. Building Global Societies on Collective Intelligence: Challenges and Opportunities. Available at SSRN 3963195 (2021)